\documentclass[fleqn,usenatbib]{mnras}

\usepackage{newtxtext,newtxmath}

\usepackage[T1]{fontenc}

\DeclareRobustCommand{\VAN}[3]{#2}
\let\VANthebibliography\thebibliography
\def\thebibliography{\DeclareRobustCommand{\VAN}[3]{##3}\VANthebibliography}

\usepackage{graphicx}	
\usepackage{amsmath}	
\usepackage{bm}
\usepackage{subcaption}
\usepackage{framed}
\usepackage{enumitem}

\title[Phase-Space DTFE]{Phase-Space Delaunay Tessellation Field Estimator}

\author[J. Feldbrugge]{
Job Feldbrugge,$^{1}$\thanks{E-mail: job.feldbrugge@ed.ac.uk}
\\
$^{1}$University of Edinburgh, Higgs Centre for Theoretical Physics, James Clerk Maxwell Building, Edinburgh EH9 3FD, UK
}

\date{Accepted XXX. Received YYY; in original form ZZZ}

\pubyear{2024}

\begin{document}
\label{firstpage}
\pagerange{\pageref{firstpage}--\pageref{lastpage}}
\maketitle

\begin{abstract}
    The reconstruction of density and velocity fields is of central importance to the interpretation of $N$-body simulations. We propose a phase-space extension of the Delaunay tessellation field estimator (DTFE) that tracks the dark matter fluid in phase-space. The new reconstruction scheme removes several artifacts from the conventional DTFE in multi-stream regions, while preserving the adaptive resolution in high-density regions and yielding continuous fields. The estimator also removes tessellation artifacts of a previously proposed phase-space reconstruction scheme. 
\end{abstract}


\section{Introduction}\label{sec:intro}
Studies of cosmic structure formation, galaxy formation, the dynamics of accretion disks, the formation of stars and planetary systems often rely on $N$-body simulations  \citep[cf.\,][]{Peebles:1971, Bertschinger:1998, Springel:2005}. An $N$-body simulation models the evolution of a physical system through the evolution of a set of particles. Physical properties involving the underlying density and velocity fields are then reconstructed from the locations and velocities of the $N$-body particles. The most well-known density reconstruction methods are the grid-based particle-in-cells (PIC) \citep[see, \emph{e.g.},][and references therein]{Harlow:1955, Harlow:1976, Harlow:1988} and the smoothed particle hydrodynamics (SPH) methods \citep[\emph{e.g.},][]{Gingold:1977, Lucy:1977}. Recently, more sophisticated density and velocity reconstruction schemes were proposed. 

The Delaunay tessellation field estimator (DTFE) \citep{Pelupessy:2003, Schaap:2007, Weygaert:2009} uses a Delaunay tessellation of the particles to adaptively increase the resolution of the density and velocity estimator in regions with many points. An alternative tessellation-based estimator was proposed by \cite{Neyrinck:2008}. The DTFE method has in the last decade become an integral part of the Disperse \citep{Sousbie:2011a, Sousbie:2011b} and NEXUS \citep{Cautun:2013, Cautun:2015} classifiers of the large-scale structure. 

The geometric pattern of the large-scale cosmic matter distribution can be understood in terms of the folding of the dark matter sheet in phase-space \citep{Arnold:1982a,Arnold:1982b, Arnold:1984, Shandarin:2012, Neyrinck:2012, Falck:2012}. Using this idea, \cite{Shandarin:2011} and \cite{Abel:2012} developed a Lagrangian density estimator that traces the evolution of the dark matter sheet in phase-space and takes into account the multi-stream nature of the cosmic web. This phase-space (PS) method is particularly powerful when the $N$-body simulation resolves the fluid in phase-space, such as in the large-scale structure at large scales, and removes several DTFE artifacts present in multi-stream regions. However, the PS reconstruction is not continuous and shows tessellation artifacts, which are particularly apparent in the single-stream regions. 

In this paper, we combine the best of the two approaches and propose the \textit{phase-space Delaunay tessellation field estimator} (PS-DTFE). This method produces a continuous reconstruction of the density and velocity fields of $N$-body simulations that removes several artifacts associated to the PS and DTFE method. Accompanying this paper, we publish the two- and three-dimensional Python implementation of the PS-DTFE\footnote{\url{www.github.com/jfeldbrugge/PS-DTFE}}.

In the following, we first briefly recap and analyze the particle-in-cells (section \ref{sec:PIC}) and smooth particle hydrodynamics (section \ref{sec:SPH}) density reconstruction schemes. We then discuss the key points of the Delaunay tessellation field estimator (section \ref{sec:DTFE}) and the phase-space density estimators (section \ref{sec:PS}). In section \ref{sec:PSDTFE}, we combine the two methods to construct the phase-space Delaunay tessellation field estimator. We discuss the implementations of this new reconstruction scheme (section \ref{sec:Implementation}) and compare the resulting density field with those constructed by the above-mentioned schemes (section \ref{sec:Comparison}).

\bigskip

\textit{Notation:} Let the $d$-dimensional $N$-body simulation consist of a set of $N$ particles located at $\bm{x}_i \in \mathbb{R}^d$ with velocities $\bm{v}_i \in \mathbb{R}^d$ (in Eulerian space) that evolved from the initial positions $\bm{q}_i \in \mathbb{R}^d$ (in Lagrangian space) with masses $m_i \in \mathbb{R}_{>0}$, for $i=1,\dots,N$. In cosmology, it is often convenient to consider these particles in a $d$-dimensional box with periodic boundary conditions. 

\begin{figure*}
    \begin{subfigure}[b]{0.32\textwidth}
        \includegraphics[width=\textwidth]{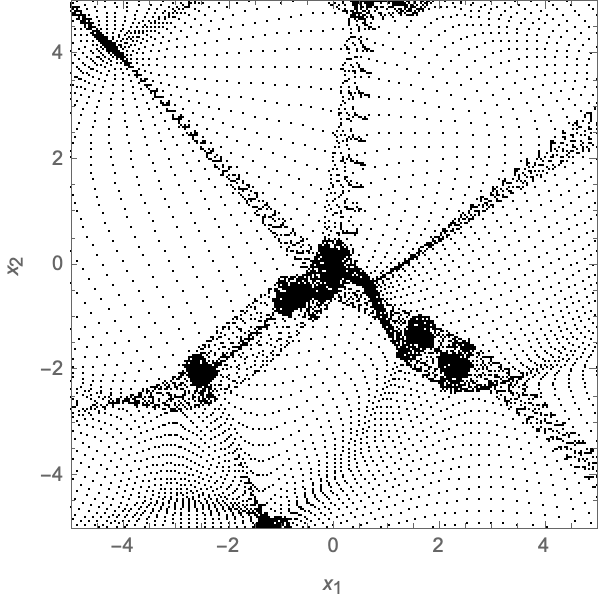}
    \end{subfigure}
    ~
    \begin{subfigure}[b]{0.32\textwidth}
        \includegraphics[width=\textwidth]{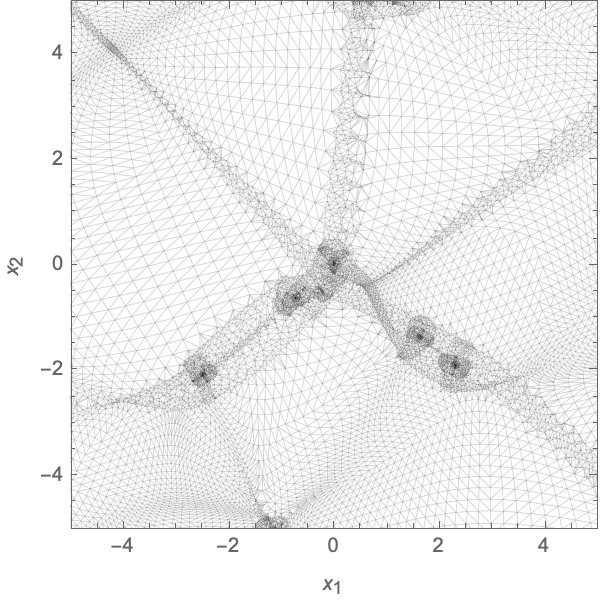}
    \end{subfigure}
    ~
    \begin{subfigure}[b]{0.32\textwidth}
        \includegraphics[width=\textwidth]{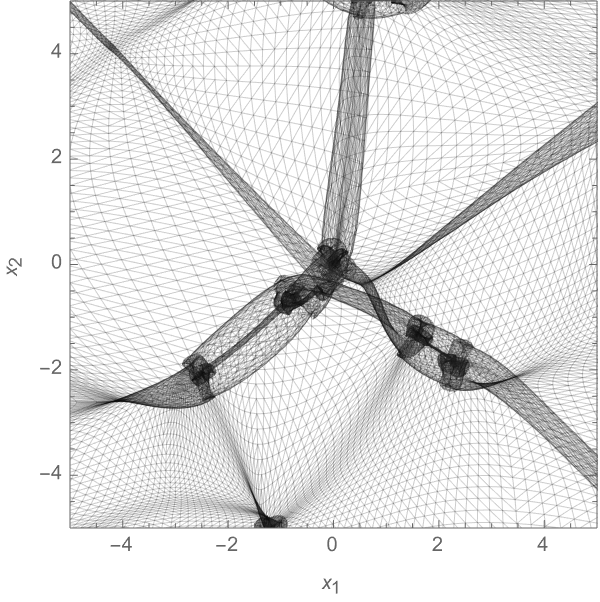}
    \end{subfigure}
    \caption{$N$-body simulation visualized in terms of the $N$-body particles (left), the Delaunay tessellation of the $N$-body particles (center), and the phase-space sheet (right).}\label{fig:Nbody}
\end{figure*}

\textit{Simulation:} The density reconstruction schemes are illustrated for a two-dimensional dark matter-only $N$-body simulation (see the left panel of fig.\ \ref{fig:Nbody}) with $N=256^2$ in a periodic box with sides $L=25$ evolving in an expanding Einstein-de Sitter universe with Gaussian initial conditions starting from a regular lattice. The particles have the same mass, \textit{i.e.}, $m_i=m$. The simulation is performed using a 2D particle-mesh $N$-body code \citep{Johan:2020}. The details of the simulation and units are not of crucial importance in this paper. The estimator works equally well for three-dimensional simulations.

\section{Particle-in-cells}\label{sec:PIC}
The particle-in-cells (PIC) density estimator is the simplest commonly used way to infer the density field from a set of particles \citep{Harlow:1955, Harlow:1976, Harlow:1988}. Assuming the particles are independently sampled from the density field, we distribute the mass of all particles onto the nodes of uniformly placed cells and divide by the volume of the corresponding cell. As the PIC estimator is solely determined by the final positions of the $N$-body simulation, we classify it as an Eulerian estimator. A well-known PIC scheme is the cloud-in-cell method \citep{Birdsall:1969}  which smears out every particle over its four nearest neighbors, weighted by the distance to each neighbor. 

The left panel of fig.\ \ref{fig:Eulerian} illustrates the cloud-in-cell density estimator for the two-dimensional $N$-body simulation discussed in section \ref{sec:intro}. The density estimator is implemented as a single for loop over the particles. Though its implementation is very efficient \citep[see][for an implementation]{Johan:2020}, the estimator generally suffers from shotnoise in void regions and does not resolve the finest features of the cosmic web. Note in particular the artifacts in the filaments, where it is clear that the particles are not independent. The resolution of the PIC routine is limited by the number of points in the simulation and effectively averages the density in the multi-stream regions.

\section{Smoothed Particle Hydrodynamics}\label{sec:SPH}
Particle-in-cells schemes rely on partitioning the cosmological volume into a set of cells leading to a uniform resolution, irrespective of the clustering of the particles. Smoothed particle hydrodynamics (SPH) density estimators \citep{Gingold:1977, Lucy:1977} are not cell-based and allow for increased resolution in high-density regions. The density consists of a convolution of the discrete particle distribution with a user-specified kernel function $W$. That is, the density at the point $\bm{x}$ is given by the sum
\begin{align}
    \rho_{SPH}(\bm{x}) = \sum_{i=1}^N m_i W(\bm{x}-\bm{x}_i,h_i)\,,
\end{align}
with the smoothing length of the kernel of the $i$th particle $h_i$. The kernel $W$ is chosen such that it is normalized to unity,
\begin{align}
    \int W(\bm{x},h)\mathrm{d}\bm{x} = 1\,,
\end{align}
to conserve the total mass of the simulation, $\int \rho_{SPH}(\bm{x})\mathrm{d}\bm{x} = \sum_{i=1}^N m_i$. The kernel satisfies the limit
\begin{align}
    \lim_{h \to 0} W(\bm{x},h) = \delta_D(\bm{x})\,.
\end{align}
Often, the kernel $W$ is assumed to be spherically symmetric $W(\bm{x},h) = W(\|\bm{x}\|,h)$. Commonly used kernels are the Gaussian kernel,
\begin{align}
    W(r,h) = \frac{1}{(2\pi h^2)^{d/2}}e^{-\frac{r^2}{2 h^2}}\,,
\end{align}
and the cubic spline kernel,
\begin{align}
    W(r,h) = \frac{\mathcal{N}}{h^d} \begin{cases}
        1-\frac{3}{2} q^2 + \frac{3}{4} q^3 & \text{when } 0 \leq q \leq 1\,,\\
        \frac{1}{4} (2 -q)^3 & \text{when } 1 \leq q \leq 2\,,\\
        0 & \text{otherwise}\,.
    \end{cases}
\end{align}
with the paremeter $q=r/h$ and the normalization factor $\mathcal{N}$ being $2/3$, $10/7\pi$ and $1/\pi$ in one, two and three dimensions, respectively  \citep[see][]{Monaghan:1992}. The two kernels yield comparable results, but the cubic spline kernel can be implemented more efficiently, as it has compact support. Originally, SPH density reconstruction was performed for a uniform smoothing length $h_i=h$ for $i=1,\dots,N$. The resolution of the density field dramatically improves when adaptively varying the smoothing lengths $h_i$ as a function of the clustering of the particles \citep{Hernquist:1989}. The SPH estimator too can be classified as an Eulerian estimator.

\begin{figure*}
    \centering
    \begin{subfigure}[b]{0.32\textwidth}
        \includegraphics[width=\textwidth]{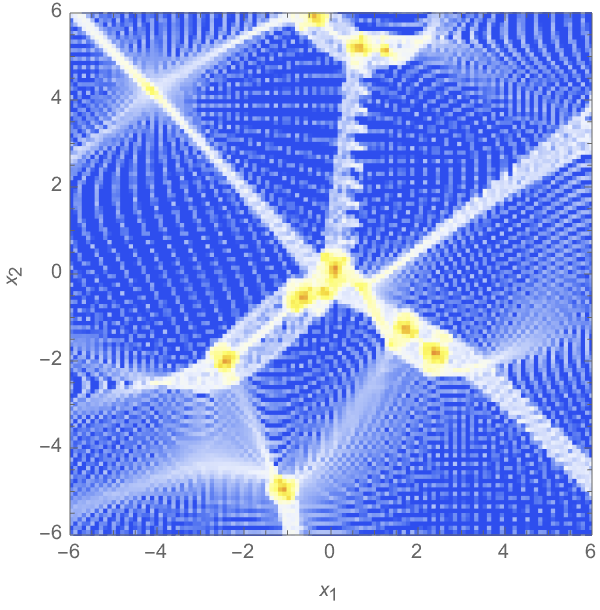}
    \end{subfigure}
    ~
    \begin{subfigure}[b]{0.32\textwidth}
        \includegraphics[width=\textwidth]{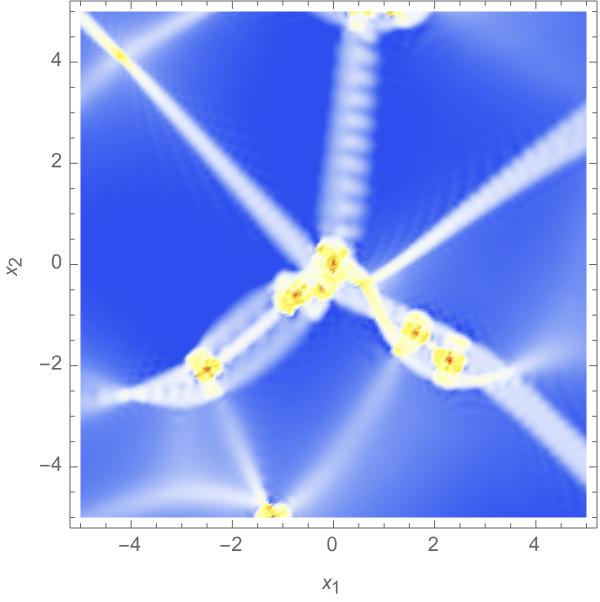}
    \end{subfigure}
    ~
    \begin{subfigure}[b]{0.32\textwidth}
        \includegraphics[width=\textwidth]{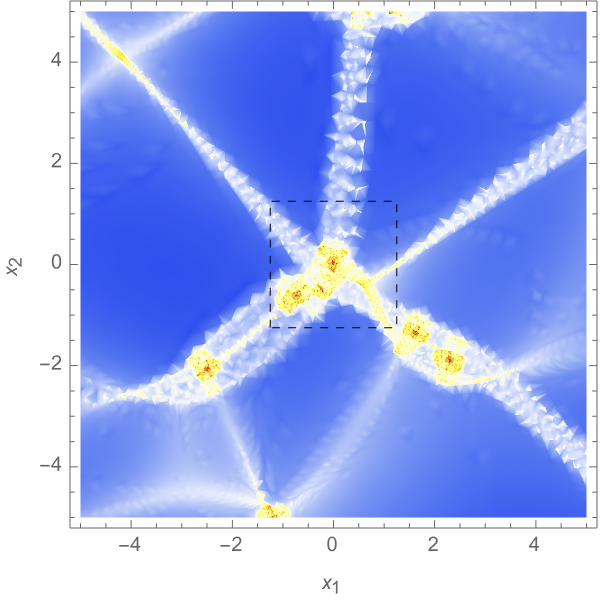}
    \end{subfigure}
    \caption{Eulerian space density field estimators on a log scale, $\log(\rho/\bar{\rho})$. The particle-in-cells density estimator (left), the smooth particle hydrodynamics density estimator (center), and the Delaunay tessellation field estimator (right).}\label{fig:Eulerian}
\end{figure*}

We illustrate the SPH density estimator with a cubic spline kernel, where $h_i$ is chosen such that the $40$ nearest neighbors lay within a distance $2h_i$, in the central panel of fig.\ \ref{fig:Eulerian}. The SPH scheme does a better job in the voids than the PIC scheme. However, the SPH density field does not capture the small-scale features of the cosmic web very well in the multi-stream regions. In particular, the SPH density field is more blurred than the PIC density estimator.

\section{Delaunay Tessellation Field Estimator}\label{sec:DTFE}
Like the SPH estimator, the Delaunay tessellation field estimator  \citep[DTFE; ][]{Pelupessy:2003, Schaap:2007, Weygaert:2009} does not rely on a uniform mesh and adapts the resolution to the clustering of particles. Yet, the DTFE scheme uses a very different approach. Instead of selecting a kernel and a procedure to determine the smoothing lengths, the DTFE scheme bases the density and velocity reconstruction on the Delaunay tessellation of the particles. The central panel of fig.\ \ref{fig:Nbody} shows a Delaunay tessellation. In regions with many particles, the simplices are smaller leading to a higher resolution of the inferred density and velocity fields. 

The DTFE scheme is defined as follows. After constructing the Delaunay tessellation, which consists of $N_T$ simplices (triangles in 2D and tetrahedra in 3D), we associate a to-be-determined density $\rho_i$ with each vertex $\bm{x}_i$. The density in a point $\bm{x}$ in a simplex $D$, spanned by $d+1$ points $\bm{x}_{l_0},\dots,\bm{x}_{l_d}$, is defined as the linear interpolation in the simplex
\begin{align}
    \rho_{DTFE}(\bm{x}) = \rho_{l_0} + [\nabla \rho] (\bm{x} - \bm{x}_{l_0})\,,
\end{align}
where the gradient $\nabla \rho \in \mathbb{R}^d$ is a vector associated to the simplex $D$ defined by
\begin{align}
    \nabla \rho = \begin{pmatrix}
        \bm{x}_{l_1} - \bm{x}_{l_0}\\
        \vdots\\
        \bm{x}_{l_d} - \bm{x}_{l_0}\\
    \end{pmatrix}^{-1}  
    \begin{pmatrix}
        \rho_{l_1}-\rho_{l_0}\\
        \vdots\\
        \rho_{l_d}-\rho_{l_0}
    \end{pmatrix}\,.\label{eq:rho_grad}
\end{align}
The integral of the density over the simplex 
\begin{align}
    \int_D \rho_{DTFE}(\bm{x})\mathrm{d}\bm{x} = \frac{V(D)}{d+1} \sum_{i \in D} \rho_i
\end{align}
is proportional to the volume of the simplex 
\begin{align} 
    V(D) = \text{abs}\left[\frac{1}{d!} \begin{vmatrix} \bm{x}_{l_1} - \bm{x}_{l_0} \\ \vdots \\ \bm{x}_{l_d} - \bm{x}_{l_0} \end{vmatrix}\right]\,,
\end{align} 
and the sum of the densities in the associated vertices $\rho_{l_0} + \dots + \rho_{l_d}$. The integral over the cosmological volume can be written as
\begin{align}
    \int \rho_{DTFE}(\bm{x})\mathrm{d}\bm{x}  &= \sum_{i=1}^{N_T} \int_{D_i} \rho_{DTFE}(\bm{x})\mathrm{d}\bm{x}\\
    &= \frac{1}{d+1} \sum_{i=1}^{N_T}V(D_i) \sum_{j \in D_i}\rho_j\,,
\end{align}
since the Delaunay tessellation partitions the cosmological volume. The first sum runs over the simplices of the tessellation. The second sum runs over the vertices of a given simplex. Noting that the to-be-determined density at a vertex $\rho_i$ is included in the sum for each simplex of which $\bm{x}_i$ is a vertex, we can rewrite this sum over the simplices as a sum over the points in the $N$-body simulation, 
\begin{align}
    \int \rho_{DTFE}(\bm{x})\mathrm{d}\bm{x}  &=  \sum_{i=1}^N \frac{\rho_i V(W_i)}{d+1}\,,\label{eq:1}
\end{align}
with $V(W_i)$ the volume of the star (or umbrella) $W_i$ of $\bm{x}_i$, defined as the union of simplices that include the vertex $\bm{x}_i$,
\begin{align}
    W_i = \bigcup_{D \text{ for which } \bm{x}_i \in D} D\,.
\end{align}
Finally, equation \eqref{eq:1} suggests a natural choice for the density $\rho_i$ associated to the vertex $\bm{x}_i$ closing the argument. When we let the density $\rho_i$ be the fraction
\begin{align}
    \rho_i = \frac{(d+1)m_i}{V(W_i)}\,,\label{eq:est}
\end{align}
the density field preserves the total mass of the $N$-body simulation, \textit{i.e.}, the integral over density coincides with the mass of the particles, $\int \rho_{DTFE}(\bm{x})\, \mathrm{d}\bm{x} = \sum_{i=1}^N m_i$. The DTFE method is very successful in cosmological simulations, in particular in the single-stream regions, as it implicitly uses the fact that the initial positions of the $N$-body particles homogeneously filled the space (corresponding to near homogeneous initial conditions). The particles are not independently distributed and can be interpreted as tracers of the evolving dark matter sheet, as we see from the occurrence of the volume $V(W_i)$ in equation \eqref{eq:est}. This is a fundamental difference between the PIC and SPH methods on the one hand and tessellation-based methods on the other hand. As the final positions of the $N$-body simulation fully determine the DTFE density, we classify it as a Eulerian estimator. 

The velocity field is also reconstructed using the Delaunay tessellation of the positions in Eulerian space. It can be reconstructed as a linear interpolation in the associated simplex, \textit{i.e.}, 
\begin{align}
    \bm{v}_{DTFE}(\bm{x}) = \bm{v}_{l_0} + [\nabla \bm{v}](\bm{x}-\bm{x}_{l_0})
\end{align}
using the matrix 
\begin{align}
    \nabla \bm{v} = \begin{pmatrix}
        \bm{x}_{l_1} - \bm{x}_{l_0}\\
        \vdots\\
        \bm{x}_{l_d} - \bm{x}_{l_0}\\
    \end{pmatrix}^{-1}  
    \begin{pmatrix}
        \bm{v}_{l_1}-\bm{v}_{l_0}\\
        \vdots\\
        \bm{v}_{l_d}-\bm{v}_{l_0}
    \end{pmatrix}\,,\label{eq:v_grad}
\end{align}
associated with the simplices $D$.

For the DTFE scheme, we first evaluate the Delaunay tessellation of the $N$-body particles and the volumes of the simplices. Subsequently, we evaluate the density estimate at the vertices $\rho_i$ using the volume of the star of $\bm{x}_i$. Given a point $\bm{x}$ in Eulerian space, we look up its associated simplex and evaluate the density and velocity field with a linear interpolation.

We illustrate the DTFE density reconstruction of the two-dimensional $N$-body simulation in the right panel of fig.\ \ref{fig:Eulerian}. Though the DTFE works very well in single-stream regions, it fails to capture the density accurately in multi-stream regions. In a multi-stream region, the volume of the Delaunay cells can get extremely small, as the Delaunay tessellation does not differentiate $N$-body particles from different streams, \textit{i.e.}, $d+1$ particles on two different streams can yield a very small Delaunay cell that does not reflect the deformation of the dark matter sheet, yielding a spike in the density field away from the shell-crossing regions. These artifacts also influence the velocity reconstruction. Recently, \cite{AragonCalvo:2021} proposed a stochastic extension of the DTFE routine, in which the particle positions are randomly perturbed and the density is defined as the average over an ensemble of density fields. Though this density estimator is not only continuous but also smooth, it suffers from the same artifacts in the multi-stream regions. The SPH and DTFE density estimators are extensively analyzed and compared by \cite{Pelupessy:2003}.

The DTFE density field is at the moment often used in the Disperse \citep{Sousbie:2011a, Sousbie:2011b} and NEXUS \citep{Cautun:2013, Cautun:2015} classification of the cosmic web into voids, walls, filaments and clusters. The artifacts of the DTFE density field in the multi-stream regions are bound to influence the classifiers in non-trivial ways.

\begin{figure*}
    \centering
    \begin{subfigure}[b]{0.49\textwidth}
        \includegraphics[width=\textwidth]{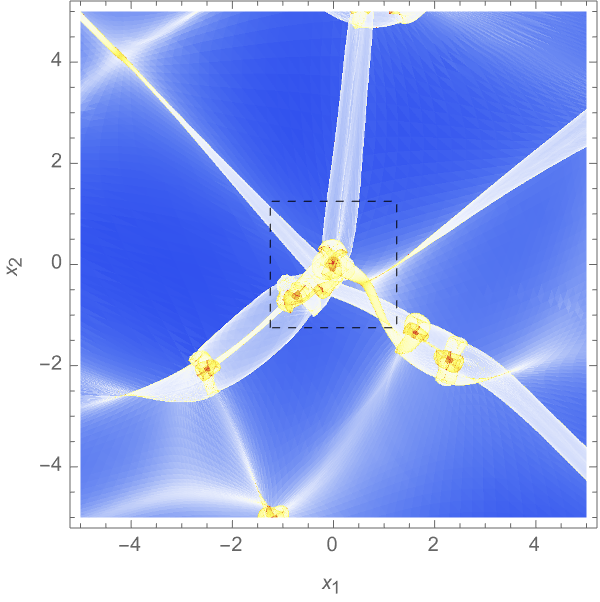}
    \end{subfigure}
    ~
    \begin{subfigure}[b]{0.49\textwidth}
        \includegraphics[width=\textwidth]{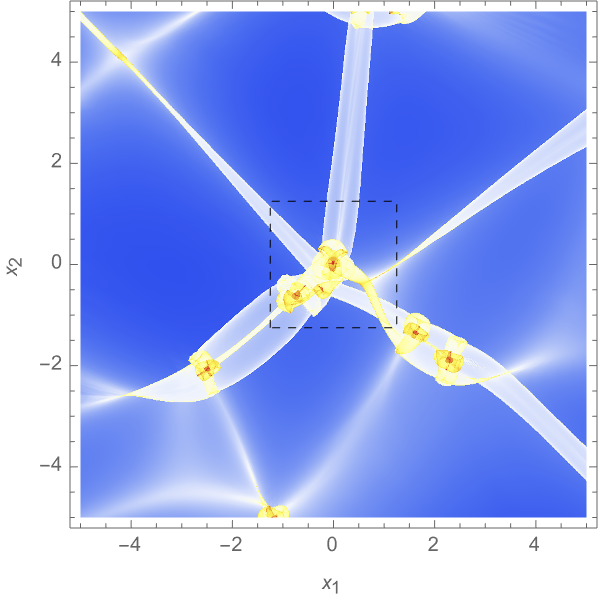}
    \end{subfigure}
    \caption{Phase-space density field estimators on a log scale, $\log(\rho/\bar{\rho})$. The phase-space density estimator (left) and the phase-space Delaunay tessellation field estimator (right).}\label{fig:Phase-Space}
\end{figure*}

\section{Phase-space Density Estimator}\label{sec:PS}
In Lagrangian fluid dynamics, the evolution of the fluid is described by the Lagrangian map 
\begin{align}
    \bm{x}_t(\bm{q}) = \bm{q} + \bm{s}_t(\bm{q})\,,
\end{align}
describing the evolution of the mass elements, traced by $N$-body particles, starting at $\bm{q}$ and displaced by $\bm{s}_t$ at time $t$. The mass elements conserve mass and the density is defined as a change of coordinates 
\begin{align}
    \rho(\bm{x}) = \sum_{\bm{q} \in \bm{x}_t^{-1}(\bm{x})} \frac{\bar{\rho}}{|\det \nabla \bm{x}_t(\bm{q})|}\,,\label{eq:LagrangianDensity}
\end{align}
where the sum ranges over the initial positions $\bm{q}$ that reach the point $\bm{x}$ in time $t$, \textit{i.e.}, $\bm{x}_t^{-1}(\bm{x})=\{\bm{q}\,|\, \bm{x}_t(\bm{q})=\bm{x}\}$.

Based on these insights, \cite{Shandarin:2011} and \cite{Abel:2012} propose a phase-space density estimator that traces the dark matter sheet. The method starts with a tessellation of the initial positions of the $N$-body particles. By freezing the tessellation and tracing the evolution of the simplices, we obtain a tessellation of the dark matter sheet in phase-space (see the right panel of fig.\ \ref{fig:Nbody}). In principle, the same phase-space matter sheet can be reconstructed from the position and velocity information at the final time. 

Assuming the $N$-body particles to have a uniform mass $m_i=m$ and start on a regular grid $\{\bm{q}_i\}$, we associate the density estimate
\begin{align}
    \rho_D = \frac{m/d!}{V(D)}
\end{align} 
to each simplex $D$ in the tessellation in Eulerian space. The phase-space density in a general point $\bm{x}$ is defined as the sum of the densities associated with the simplices that include the point $\bm{x}$, \textit{i.e.},
\begin{align}
    \rho_{PS}(\bm{x}) = \sum_{D\text{ for which } \bm{x} \in D} \rho_{D}\,.
\end{align}

We illustrate this density estimator for the two-dimensional $N$-body simulation (see the left panel of fig.\ \ref{fig:Phase-Space}). This phase-space method has the advantage that it does not mistake particles on different streams as neighbors and cleanly resolves the density in the multi-stream regions of the $N$-body simulation. Several artifacts present in the Eulerian density estimators are removed. However, the scheme is discontinuous at the edges of the simplices \citep[though smooth generalizations are discussed;][]{Abel:2012}, as we can observe in single-stream regions of the cosmic web and also in multi-stream regions, but to a lesser extent. The phase-space estimator outperforms the DTFE estimator in the multi-stream regions. On the other hand, the DTFE estimator performs better than the phase-space estimator in the single-stream regions. Note that this phase-space method can only be applied when we have access to the initial position of each particle in the distribution and the dark matter sheet is resolved by the $N$-body simulation.

\begin{figure*}
    \centering
    \begin{subfigure}[b]{0.49\textwidth}
        \includegraphics[width=\textwidth]{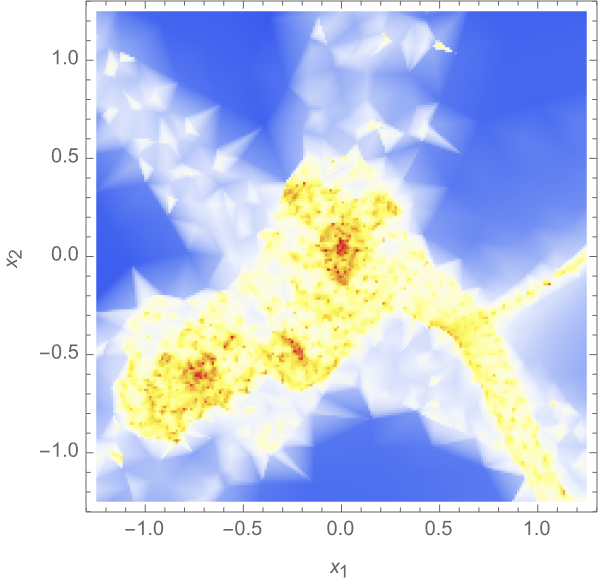}
    \end{subfigure}
    ~
    \begin{subfigure}[b]{0.49\textwidth}
        \includegraphics[width=\textwidth]{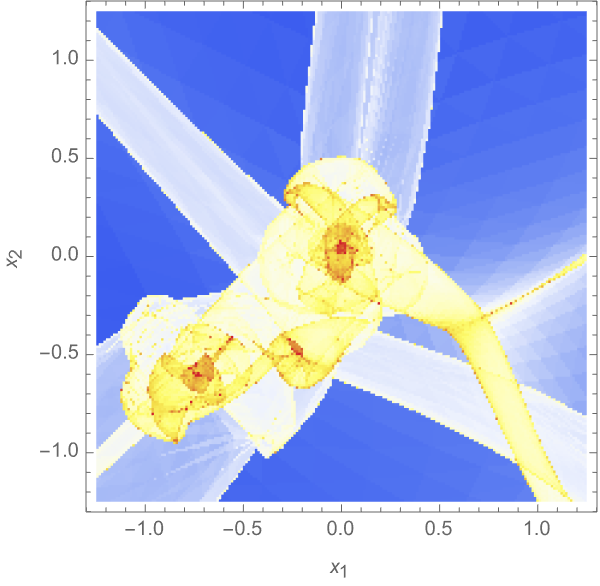}
    \end{subfigure}\\
    \begin{subfigure}[b]{0.49\textwidth}
        \includegraphics[width=\textwidth]{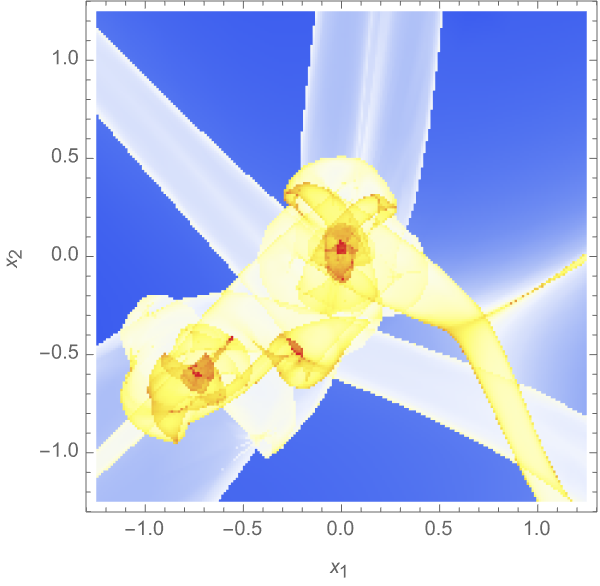}
    \end{subfigure}
    ~
    \begin{subfigure}[b]{0.49\textwidth}
        \includegraphics[width=\textwidth]{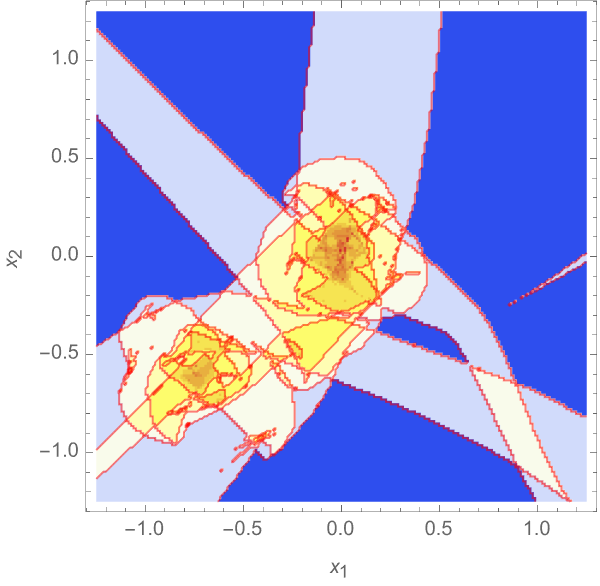}
    \end{subfigure}
    \caption{A close-up view of the tessellation density field estimators on a log scale, $\log(\rho/\bar{\rho})$ and the number of streams (in a box corresponding to the dashed boxes in figs.\,\ref{fig:Eulerian} and \ref{fig:Phase-Space}). The DTFE (upper left), the PS (upper right), and the PS-DTFE estimator (lower left). The number of streams on a logarithmic scale with the fold caustic (red curve) is plotted in the lower right panel.}\label{fig:Closeup}
\end{figure*}

\begin{figure*}
    \centering
    \includegraphics[width=\textwidth]{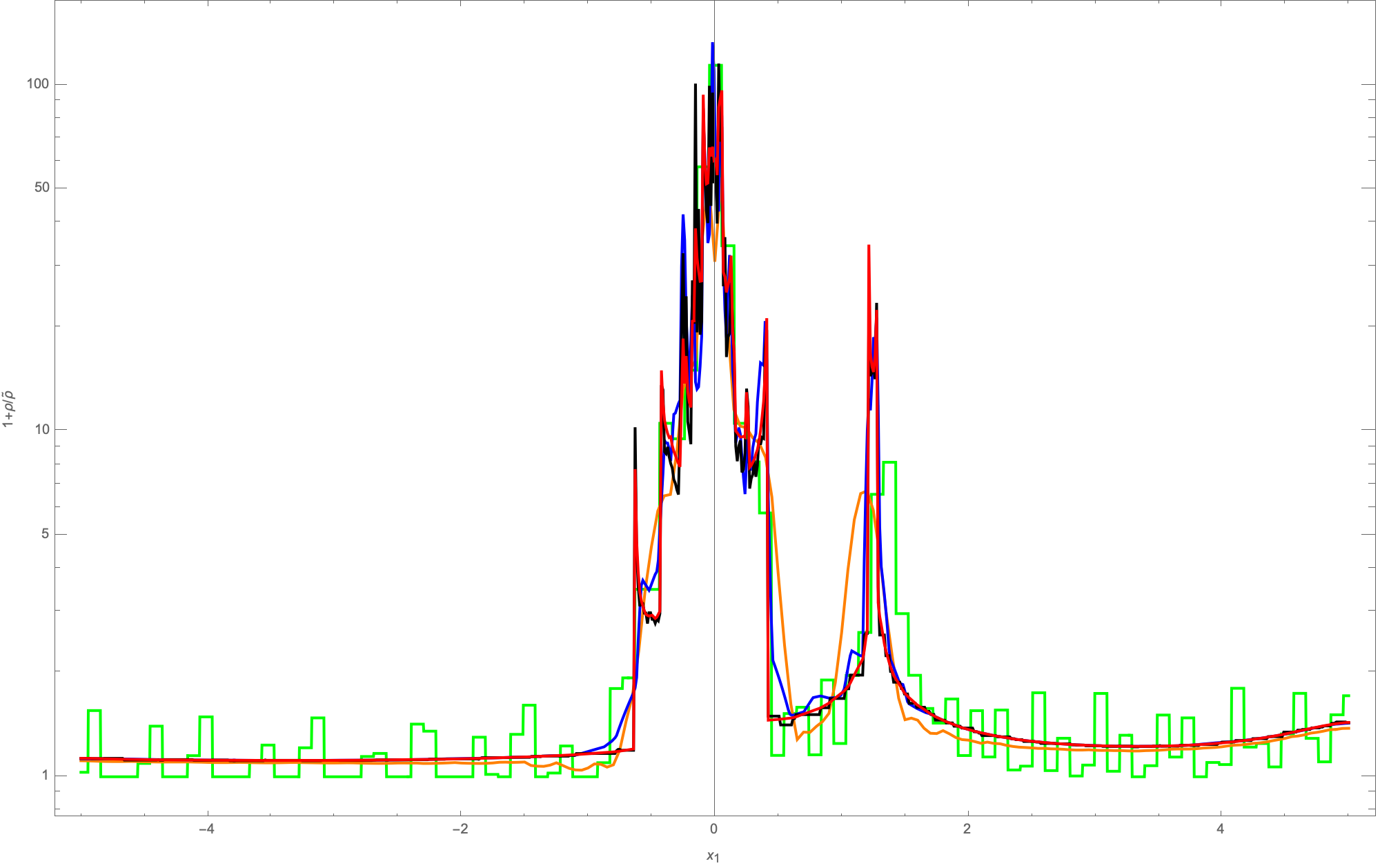}
    \caption{The log density field along the line $x_2=0$. The particle-in-cells estimator (green), the smooth particle hydrodynamics estimator (orange), the Delaunay tessellation field estimator (blue), the phase-space estimator (black) and the phase-space Delaunay tessellation field estimator (red).}\label{fig:comparison}
\end{figure*}

\section{Phase-space Delaunay Tessellation Field Estimator}\label{sec:PSDTFE}
By combining the DTFE and phase-space estimators, in this paper, we extend the DTFE method and obtain a new reconstruction scheme that smoothly reconstructs the density and velocity fields tracing the dark matter sheet in phase-space, adaptively increasing the resolution in high-density regions. We call this the \textit{phase-space Delaunay tessellation field estimator} (PS-DTFE).

For a $N$-body simulation starting with a regular grid $\{\bm{q}_i\}$ (and potentially variable particle mass), we first evaluate the Delaunay tessellation of the initial positions of the particles. To obtain a unique tessellation, we need to randomly perturb the initial positions slightly. Alternatively, we can evaluate the Delaunay tessellation of the particles early in their evolution, before the first shell-crossing event. Fixing the tessellation and mapping the particles to their final positions $\{\bm{x}_i\}$, we obtain a tessellation of the dark matter sheet in Eulerian space, like we saw in the phase-space density reconstruction method (see the right panel of fig.\ \ref{fig:Nbody}). Next, for each vertex $\bm{x}_i$, we estimate its density as the fraction of the mass of the particle divided by its star in Eulerian space,
\begin{align}
    \rho_i = \frac{(d+1)m_i }{V(W_i)}\,.\label{eq:regularGrid}
\end{align}
Finally, for a general point in Eulerian space $\bm{x}$, we evaluate the simplices in which it lies and sum over the linear interpolations of the density in the simplices, \textit{i.e.},
\begin{align}
    \rho_{PS\text{-}DTFE}(\bm{x}) = \sum_{D\text{ for which } \bm{x} \in D} \rho_{l_0} + [\nabla \rho] (\bm{x} - \bm{x}_{l_0})\,,
\end{align}
where the gradient $\nabla \rho$ associated to the simplex $D$, spanned by $\bm{x}_{l_0},\dots,\bm{x}_{l_d}$, is given by equation \eqref{eq:rho_grad}. Let $N_T$ again denote the number of simplices. Like the DTFE, this density estimate conserves the mass of the $N$-body particles, as
\begin{align}
    \int \rho_{PS\text{-}DTFE}(\bm{x}) \mathrm{d}\bm{x} &= \sum_{i=1}^{N_T} \int_{D_i} \rho_{PS\text{-}DTFE}(\bm{x}) \mathrm{d}\bm{x}\\
    &= \sum_{i=1}^{N_T} V(D_i) \sum_{j \in D_i}\rho_j\\
    &=  \sum_{i=1}^N \frac{\rho_i V(W_i)}{d+1}
    = \sum_{i=1}^N m_i.
\end{align}
. 

When the initial distribution of the $N$-body particles $\{\bm{q}_i\}$ do not lay on a regular grid but the initial conditions are assumed to be close to a uniform density field \citep[for example, in the case of glass initial conditions, as proposed by][]{White:1994, Baugh:1995}, we start with a Delaunay tessellation of the initial conditions and associate the density 
\begin{align}
    \rho_i = \frac{\bar{\rho}  \, V(W_{L,i})}{V(W_i) }
\end{align}
to each vertex, with the mean cosmological density $\bar{\rho}$ and the volume of the star (umbrella) of the point $\bm{x}_i$ in Lagrangian space $V(W_{L,i})$. When the simplex $D$ is spanned by the points $\bm{x}_{l_0},\dots, \bm{x}_{l_d}$, the simplex $D_{L,i}$ is spanned by the points $\bm{q}_{l_0},\dots, \bm{q}_{l_d}$. This formula yields an adaptive continuous implementation of the Jacobian in equation \eqref{eq:LagrangianDensity} and coincides with the estimator \eqref{eq:regularGrid} for $N$-body simulations starting from a regular grid. This density estimate too conserves the total mass, 
\begin{align}
    \int \rho_{PS\text{-}DTFE}(\bm{x}) \mathrm{d}\bm{x} 
    &=  \sum_{i=1}^N \frac{\rho_i V(W_i)}{d+1}\\
    &= \bar{\rho} \sum_{i=1}^N \frac{V(W_{L,i})}{d+1}\\
    &= \bar{\rho} \sum_{i=1}^{N_T} V(D_{L,i})\,,
\end{align}
with $\sum_{i=1}^N V(D_{L,i})$ the total volume of the simulation box.

The velocity field reconstruction follows analogously. For each simplex, we evaluate the matrix $\nabla \bm{v}$, following equation \eqref{eq:v_grad}. The reconstructed velocity field is multi-valued in the multi-stream regions,
\begin{align}
    \bm{v}_{PS\text{-}DTFE}(\bm{x}) = \{ &\bm{v}_{l_0} + [\nabla \bm{v}](\bm{x}-\bm{x}_{l_0})\, |\, \nonumber \\
    &\text{ for all simplices } D \text{ that include } \bm{x}\}\,.
\end{align}

This method automatically adapts the resolution of the reconstruction, is continuous, conserves the total mass of the $N$-body simulation and simultaneously yields accurate results in the multi-stream regions. The right panel of fig.\ \ref{fig:Phase-Space} illustrates the PS-DTFE density reconstruction scheme of the two-dimensional $N$-body simulation. Comparing the PS with the PS-DTFE reconstruction, we see that the PS-DTFE no longer shows the underlying tessellation. A close look indicates that the PS-DTFE density estimator removes noise in both the single- and the multi-stream regions (see fig.\ \ref{fig:Closeup}). This can be attributed to the fact that we effectively average the density when we evaluate the volume of the star of a vertex.

\section{Implementation}\label{sec:Implementation}
The efficient implementation of any tessellation-based density and velocity field reconstruction method relies on a quick algorithm to repeatedly intersect the simplices with a general point. For the PS-DTFE estimator, we implement this search using a bounding volume hierarchy \citep[see, for example,][]{Goldsmith:1987, Ericson:2004}. By organizing the simplices in a tree, based on the subvolumes they reside in, we significantly decrease the number of simplices whose intersections need to be checked. An added benefit of the hierarchy is that it also yields the number of streams at any point in Eulerian space.

Before reconstructing the density and velocity field at a given point, we follow three steps:
\begin{tabbing}
    \,\,\,(i)\, \= we evaluate the Delaunay tessellation of the initial conditions \\
\> (or an early stage of the evolution),\\
\,(ii) \> we construct a bounding volume hierarchy of the simplices,\\
(iii) \> and we evaluate the volumes of the simplices, and compute the \\
\> density estimates at the vertices and gradients at the simplices. 
\end{tabbing}

Once these structures are constructed, we evaluate the density and velocity in a general point:
\begin{tabbing}
    \,\,(iv)\, \= we find the corresponding simplices using the bounding volume \\
\>hierarchy and a for loop over the candidate simplices,\\
    \,\,\,(v)\, \> we sum over the densities of the various streams,\\
\,(vi) \> and we evaluate the velocities of the different streams.
\end{tabbing}

In the accompanying code, we implement the DTFE, phase-space, and PS-DTFE reconstruction methods for two- and three-dimensional $N$-body simulations in Python\footnote{Code available at: \url{www.github.com/jfeldbrugge/PS-DTFE}\,.}. The Delaunay tessellations are evaluated with the \texttt{scipy} package \citep{SciPy}. To aid the application of the methods to other projects, the $N$-body simulation in the left panel of fig.\ \ref{fig:Nbody} is included along with the code.

The computational cost of the PS-DTFE scheme is comparable to the cost of the DTFE and the PS schemes. Compared to the DTFE scheme, we need to add the bounding volume hierarchy as the Delaunay tessellation of the \texttt{scipy} package already includes a quick routine to intersect the simplices with a general point. Compared to the PS scheme, the added computational costs come from the evaluation of the gradients $\nabla \rho$ and $\nabla \bm{v}$ and the linear interpolation.

\section{Comparison}\label{sec:Comparison}
We compare the reconstruction methods in the context of large-scale structure formation, noting that the density fields in figs.\ \ref{fig:Eulerian} and \ref{fig:Phase-Space} as well as the closeups in fig.\ \ref{fig:Closeup} are plotted with the same color scheme. From fig.\ \ref{fig:Eulerian}, we see that the PIC method has large artifacts in the single-stream regions. The adaptive sampling of the SPH and DTFE method leads to a significant improvement, however, there are still significant artifacts visible in the multi-stream regions. Generally, the Eulerian methods are unable to capture the fact that $N$-body particles may belong to different streams. Note that whereas the PIC and SPH schemes rely on a set of choices (the way to distribute mass over the nodes, the kernel, and smoothing lengths), the DTFE reconstruction scheme is parameter-free. The Delaunay tessellation and the linearity and continuity assumptions yield a single natural reconstruction method. From fig.\ \ref{fig:Phase-Space}, we see that both phase-space methods remove these artifacts. While the tessellation is still perceivable in the PS reconstruction, it is almost completely absent in the PS-DTFE scheme. This is particularly clear in the close-up view of the tessellation reconstruction methods in fig.\ \ref{fig:Closeup}. Note that the tessellation-based reconstruction methods are all parameter-free. All the presented methods conserve the total mass of the $N$-body particles.

The PIC method is the cheapest to evaluate. Yet, when comparing the density reconstructions along the slice $x_2=0$ (see fig.\ \ref{fig:comparison}), we again observe that the PIC estimator performs the poorest, with many fluctuations in the single-stream regions. The SPH estimator performs better but is unable to capture the caustics. We might get a slightly better result when changing the rule to fix the smoothing length scale, but probably not by much. The DTFE outperforms the SPH estimator in the single-stream regions but leads to large fluctuations in the multi-stream regions. The density field near the caustics is not correctly captured by the DTFE scheme. The PS method correctly captures the density field near the caustics but is discontinuous. This property is easiest seen in the single-stream regions, but can also be observed in the high-density multi-stream regions in the form of noise. Finally, the PS-DTFE nicely combines the beneficial properties of the DTFE and the PS reconstruction routines. The density field is  smooth everywhere. The effective smoothing due to the evaluation of the volume of the star, instead of the evaluation of a single simplex, yields a smoother density curve in the multi-stream regions.

\section{Conclusion}\label{sec:Conclusion}
We propose a new density and velocity reconstruction routine, extending the Delaunay tessellation field estimator to trace the dark matter fluid in phase-space. We demonstrate that the resulting density and velocity fields are continuous and remove artifacts present in the DTFE scheme in the multi-stream regions while conserving the total mass of the $N$-body particles. Moreover, the phase-space DTFE method also removes some artifacts present in the earlier proposed phase-space reconstruction method. A Python implementation for two- and three-dimensional $N$-body simulations is provided (see section \ref{sec:Implementation}). 

In the PS-DTFE scheme, we compute the Delaunay tessellation of the $N$-body particles at an early time and push this tessellation to the late-time universe. While this is not a problem when we work with $N$-body simulation, a method to construct the dark matter sheet directly from the position and velocity data is required to apply these methods to cosmological redshift surveys.

Given that the PS-DTFE method yields a significant improvement over the previously proposed DTFE method, we will in an upcoming paper explore whether the NEXUS and Disperse cosmic web classifiers can be improved with this new density estimator. 

Finally, in the future, it would be beneficial to implement the phase-space Delaunay tessellation field estimator in a fully compiled language, such as C++, to improve the speed of the evaluation of the density and velocity fields. 

\section*{Acknowledgements}
I thank Rien van de Weygaert for introducing me to tessellation-based reconstruction methods and Sergei Shandarin for introducing me to the phase-space perspective on the cosmic web. Finally, I thank Nynke Niezink for many discussions and for proofreading this manuscript. The work is supported by the STFC Consolidated Grant ‘Particle Physics at the Higgs Centre’ and by a Higgs Fellowship at the University of Edinburgh.

\section*{Data Availability}
No new data were generated or analyzed in support of this research. The implementation of the tessellation-based reconstruction methods and the simulation used to compare the different methods are available at \url{www.github.com/jfeldbrugge/PS-DTFE}.

\bibliographystyle{mnras}
\bibliography{Library} 


\appendix

\bsp	
\label{lastpage}
\end{document}